\documentclass[10pt,conference]{IEEEtran}
\usepackage[T1]{fontenc}
\usepackage[english]{babel}
\usepackage{amsmath,amssymb}
\usepackage{graphicx}
\usepackage{booktabs}
\usepackage{cite}
\usepackage{url}
\usepackage[hidelinks]{hyperref}
\hypersetup{pdftitle={Toward Operationalizing Rasmussen in Evolving Software Systems},pdfauthor={Anatoly A. Krasnovsky}}

\newcommand{\cl}{\mathcal{C}}
\title{Toward Operationalizing Rasmussen\\in Evolving Software Systems}
\author{\IEEEauthorblockN{Anatoly A. Krasnovsky}
\IEEEauthorblockA{Innopolis University and MB3R Lab\\
Innopolis, Russia}}
\begin{document}
\maketitle

\begin{abstract}
Rasmussen's qualitative drift model explains how successive, locally successful adaptations under efficiency pressures can move a system toward unacceptable operation. Brooker and Hochstein applied this perspective conceptually to software operations without specifying a quantitative procedure for observing drift. We take a step in this direction and propose a runtime observer that quantitatively operationalizes the model's observable part for evolving software. The proposed observer is intended to give engineers evidence for pausing or revising adaptation before declared operating limits are crossed. It tracks how activity or resource use is distributed across components and types of work. Measurement history quantifies accumulated change and recent movement toward or away from current limits. Assuming that direction and limits remain unchanged, the proposed observer calculates the first constraint ahead and the amount of further change required to reach it. Measurements remain comparable across supported component changes, while policy revisions are distinguished from changes in execution. Preliminary runtime studies demonstrated these distinctions under annotated software changes and revised retry limits. A separate multivariate exponentially weighted moving average (MEWMA) experiment detected CPU redistribution earlier with log-ratio coordinates than with raw shares, motivating investigation of representation-dependent observation. Further work will ground constraints in service outcomes, investigate runtime-model acquisition, and test whether decisions informed by observed drift reduce violations at acceptable useful-work loss and intervention cost.
\end{abstract}

\section{From Qualitative Drift to Quantitative Observation}
Rasmussen described how successive, locally successful adaptations under economic and workload pressures can move a system toward unacceptable operation~\cite{rasmussen1997}. Proactive risk management requires feedback on this process~\cite{rasmussen2000}. Brooker called for feedback between software incidents~\cite{brooker2014}; Hochstein interpreted SRE through the safety boundary~\cite{hochstein2020}. Neither account specified a quantitative procedure for observing drift.

The qualitative diagram leaves telemetry variables, units of movement, and measurable boundaries unspecified. Software evolution adds a further difficulty: a component split or replacement can alter the meaning of a counter, while a policy revision can alter a diagnosis without changing execution. These changes must remain distinguishable.

Retries motivate the problem: an extra attempt can recover a failed call while adding work constrained by retry budgets~\cite{sre_overload}. The current retry fraction alone does not distinguish movement toward a limit from recovery. Reviewing successive retry-policy and deployment changes also requires knowing whether measurements still describe the same logical work.

We propose a quantitative operationalization of the observable part of Rasmussen's model to inform engineers' decisions to pause or revise adaptation before declared limits are crossed, even after locally successful changes. A retained reference and comparable telemetry define accumulated displacement and recent movement; versioned policies delimit admissibility. Component correspondence preserves comparisons through supported structural changes.

Our contribution comprises the observation model, an executable instance, and emerging evidence. Eight histories examine reconstruction under evolution and conditional movement relative to policies. A separate MEWMA experiment examines representation-dependent detection of resource-state change. Subsequent studies will evaluate operational decisions using the measured trajectory.

\begin{figure*}[t]
\centering
\includegraphics[width=\textwidth]{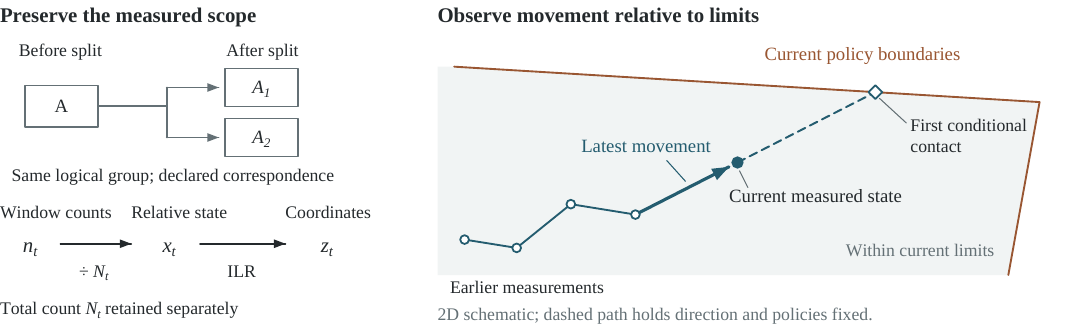}
\caption{Observation model for evolving software. Declared component correspondence preserves the logical scope across structural changes. Counts define comparable relative states while absolute volume is retained. Measured history gives the latest direction; current policies delimit admissibility. The dashed continuation reaches the diamond at conditional first contact if direction and policies stay fixed. The layout is schematic; computation retains all coordinates.}
\label{fig:model}
\end{figure*}

\section{Quantitative Observation Model}
The proposed observer turns successive telemetry windows into a trajectory whose components retain their meaning as software changes. It assesses this trajectory against current limits: execution may move the operating point, while a policy revision may move a boundary (Figure~\ref{fig:model}).

\subsection{State and geometry}
We select a logical scope and aggregate its observations through a versioned component mapping. For nonnegative counts $n=(n_1,\ldots,n_D)$ with positive total, we retain both absolute volume and relative state:
\begin{equation}
N=\sum_i n_i,\qquad x=n/N,\qquad \sum_i x_i=1.
\label{eq:state}
\end{equation}
For example, proportional growth of all counts changes $N$ while leaving $x$ unchanged. Closure is introduced by this choice of normalized observations; Rasmussen's bounded operating region alone does not imply a simplex, and an event total need not be a fixed physical budget.

Closure couples the parts, so measuring relative movement requires a choice of geometry. Compositional data analysis treats ratios as the relevant relations~\cite{aitchison1982,egozcue2019}. For positive parts, we use isometric log-ratio coordinates~\cite{ilr2003}:
\begin{equation}
z=V\log x,\qquad x=\cl\exp(V^\top z).
\label{eq:ilr}
\end{equation}
The fixed matrix $V$ satisfies $V\mathbf{1}=0$ and $VV^\top=I$; $\cl$ normalizes positive parts to unit sum. All $D-1$ coordinates are retained. The four-part instance has three ILR coordinates; Figure~\ref{fig:model} is a two-dimensional schematic, with calculations retaining all coordinates.

A retained reference records net displacement $z_t-z_{\mathrm{ref}}$. Comparable adjacent states provide the latest increment $v_t=z_t-z_{t-1}$ and its exponential smoothing~\cite{roberts1959}. These describe accumulated change and recent movement on distinct time scales. A structural update that preserves the logical scope need not erase that history.

\subsection{Movement and current constraints}
For either nonzero latest or smoothed direction $v$, we inspect the conditional continuation
\begin{equation}
x(s)=\cl\exp\!\left(V^\top[z+s\,v/\|v\|]\right),\quad s\geq0.
\label{eq:path}
\end{equation}
Holding current policies fixed, the observer reports movement toward or away from each constraint, the first constraint encountered, and the distance to it. Here $s$ is Aitchison path length. A temporal forecast would additionally require a model of how quickly and how persistently execution follows that direction.

Our runtime instance counts attempt starts within each observation window. For logical group $g$, $F_g$ counts initial attempts and $R_g$ counts all repeated attempts, including second and later starts. The groups are catalog ($c$) and currency ($u$) reads, giving $n=(F_c,R_c,F_u,R_u)$. The retry fraction $r_g=R_g/(F_g+R_g)$ is the share of attempt starts that are retries; it is not the fraction of calls that fail. Each group has policy $r_g\leq\rho_g$. Its margin $m_g=\rho_g-r_g$ admits the decomposition
\[
\Delta m_g=\Delta\rho_g-\Delta r_g
\]
when the logical scope is comparable. This separates movement of the measured state from movement of the constraint. The policy also has an exact log-ratio form, $\log(R_g/F_g)\leq\log[\rho_g/(1-\rho_g)]$, which permits analytic contact calculation along~\eqref{eq:path}. These limits define an operationally declared region; their connection to resource exhaustion or service failure remains to be established.

\subsection{Comparability during evolution}
Deployment annotations identify logical groups. We supplement service identity~\cite{otel_service} with declared predecessor records that propagate membership through supported splits, replacements, and merges. The observer assesses a completed window under the policy effective at snapshot generation. Retaining the window, mapping version, and policy timing provides assessment provenance~\cite{prov2013}. A revised limit can reassess the same window without adding an execution increment.

Unresolved correspondence blocks affected assessments. Zero handling requires a modeling choice~\cite{zeros2003}: a nonempty group's local fraction remains defined at zero retries, while this instance withholds full log-ratio geometry. A gap starts a new movement segment. These availability rules distinguish unavailable observations from stability.

Two consistency requirements govern operationalization: supported repartition preserves aggregate state when counted events and logical scope remain unchanged; revised requirements can change the diagnosis without an execution increment. Correspondence and policy versions identify comparable measurements and applicable requirements, making changed assessments traceable to execution or policy. Organizational pressures remain outside this measured projection.

\begin{figure*}[t]
\centering
\includegraphics[width=\textwidth]{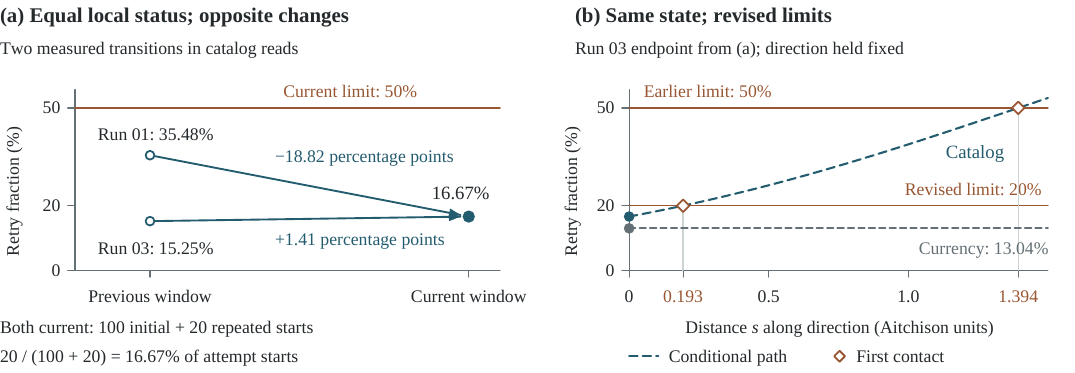}
\caption{Two distinctions recovered from runtime observations. (a) Two executions reach the same catalog retry fraction under a 50\% limit, with opposite latest changes (Run~01: windows 9--10; Run~03: 13--14). Their full compositions differ. (b) Reassessment of Run~03, window 14, holds the full state and latest direction fixed while both group limits fall from 50\% to 20\%. Both groups remain compliant; catalog's conditional first-contact distance decreases from 1.394 to 0.193. Points at $s=0$ are observed; dashed curves follow~\eqref{eq:path}. Diamonds mark computed contacts. Distance $s$ is Aitchison path length, not elapsed time.}
\label{fig:examples}
\end{figure*}

\section{Preliminary Evaluation and Emerging Results}
\subsection{Observation under evolving execution}
Eight recorded executions used real OpenTelemetry Demo catalog/currency reads~\cite{otel_demo,artifact} through an evolving forwarding/retry layer. Each contained 19 nonoverlapping ten-second windows. Scripted delivery faults followed successful upstream reads; the attempt allowance increased from two to three. Splits, replacements, a merge, and policy revisions exercised the observation model. Passive observers produced state/policy reports online. Directional analysis subsequently used saved causal prefixes.

Three variants received the same snapshots and available metadata: a fixed initial schema, updated direct fractions, and updated complete relative state. Event recounting checked available states and effective policies. Table~\ref{tab:state} reports the results. Both updated observers recovered 304 correct available assessments; their agreement attributes the improvement over the fixed schema to maintaining correspondence and policy interpretation.

\begin{table}[tb]
\centering
\caption{State/policy assessment over eight histories. Each observer had 362 group assessments; observations within a history are dependent.}
\label{tab:state}
\begin{tabular}{lrrr}
\toprule
 & Fixed & Updated & Updated\\
 & schema & fractions & relative state\\
\midrule
Available &129&304&304\\
Correct state/policy/margin &80&304&304\\
Wrong policy decisions &48&0&0\\
\bottomrule
\end{tabular}
\end{table}

The histories recorded a local benefit together with additional work: 667 calls succeeded on a third attempt after two unsuccessful attempts. Figure~\ref{fig:examples}(a) compares two executions with equal current catalog counts ($F_c=100$, $R_c=20$) and a 50\% limit. Run~01's catalog fraction decreased to 16.67\%; Run~03's increased to the same value. Both had a 33.33-percentage-point margin, but their latest changes pointed away from and toward the limit, respectively. The full compositions differed.

Figure~\ref{fig:examples}(b) then takes the final Run~03 observation from panel (a) and revises both limits from 50\% to 20\%. With the full state and latest direction unchanged, the conditional first-contact distance decreased from 1.394 to 0.193. Along this path, catalog's fraction increases from 16.67\%, while currency's stays at 13.04\%. The revised diagnosis therefore followed a policy change without deterioration in the measured execution. Across the corpus, 13 policy-only reassessments preserved state and direction, and five changed current compliance. The two panels distinguish the roles of execution history and policy history when reviewing an adaptation sequence.

Direction and policies were available for 119 of 181 report batches (65.7\%). Independent calculations agreed on all 238 latest/smoothed conditional diagnoses along the same log-linear path. The two log-odds $q_g=\log(R_g/F_g)$, proportional to two-part balances~\cite{balances2005}, reproduced policy orientations and first constraints on the common availability mask. These outputs do not require the entire composition.

\subsection{Geometry-sensitive detection with MEWMA}
NIST describes EWMA monitoring for small or gradual shifts and its multivariate extension~\cite{nist_ewma,nist_mewma}. MEWMA retains a weighted history of deviations from a reference and evaluates a covariance-adjusted statistic. Compositional MEWMA~\cite{mewma2018,nguyen2022} and CUSUM~\cite{imran2022} are established; here MEWMA supplies an experiment on representation-dependent observation in software.

A separate campaign compared raw-share and ILR observers over 250 qualified CPU-reallocation/control pairs~\cite{artifact}. The measured state comprised CPU-consumption shares of 11 components. Both variants retained all shares in ten orthonormal contrasts, $Vx$ or $V\log x$, avoiding the singular covariance of the closed eleven-part vector. Both used full covariance and the same temporal weighting. Fitting and threshold calibration used separate background histories; observer parameters were frozen before evaluation, and each execution retained its reference during gradual quota transfers.

Matched controls retained their allocation. Cycles lasted 40~s; delay counted post-intervention cycles, restricted to 12 with an undetected run assigned 12. Both original observers detected all 250 reallocation histories. ILR signaled earlier in 207 pairs, tied in 40, and later in three. Table~\ref{tab:mewma} reports the mean advance and control signals.

\begin{table}[!ht]
\centering
\caption{Separate CPU/MEWMA campaign, 250 pairs. Positive advance favors ILR. The second row reuses evaluation controls retrospectively.}
\label{tab:mewma}
\begin{tabular}{lrr}
\toprule
ILR threshold &Mean advance&Control signals\\
 & (cycles)&ILR / raw\\
\midrule
Original $h$ &1.252&9 / 6\\
Retrospective $1.10h$ &0.972&6 / 6\\
\bottomrule
\end{tabular}
\end{table}

The original 1.252-cycle advantage is a 50.08~s difference in restricted mean detection delay. A stricter ILR threshold retained an advantage at equal observed control counts, which do not establish equal population false-alarm rates. Representation, covariance estimation, and calibration were not causally separated. The result supports investigating geometry for drift detection; it does not measure lead time to a service violation or the benefit of intervention.

\section{Positioning and Scope}
Reviewing the cumulative effects of locally successful adaptations requires relating comparable observations to the operating constraints currently in force. Architectural and runtime models support adaptation~\cite{rainbow2004,blair2009}; temporal queries, provenance, and dynamic assurance cases retain evidence across execution and change~\cite{runtime_history2022,runtime_provenance2020,entrust2018}. Our contribution specifies how comparable observations form a trajectory relative to current constraints. Trace-based discovery supplies structural evidence~\cite{discovery2026}; preserving logical scope additionally requires correspondence across versions. Our instance used declared annotations and predecessor records.

Interpreting movement against limits connects to runtime verification and quantitative robustness, which evaluate execution against specifications~\cite{leucker2009,fainekos2009}, and to ecological interface design, which emphasizes visible work-domain constraints~\cite{vicente1992}. Rasmussen-inspired studies of lost buffering and system dynamics provide context outside software engineering~\cite{cook2005,morrison2022}. The proposed observer gives Rasmussen's notion of drift a measurable interpretation through accumulated displacement and conditional movement relative to declared constraints. Organizational pressures remain outside the measured projection.

Geometry is a methodological choice within this operationalization. CoDA has been applied to software-effort allocation~\cite{effort2015}, and anomaly detection on the simplex has been studied under different geometries~\cite{legaria2023}. Our retry study found agreement between reduced log-odds and full-state calculations on movement toward limits and the first constraint encountered, along the same path on jointly available observations. The CPU experiment observed different MEWMA crossing times, with representation, covariance estimation, and calibration not causally separated. These findings motivate evaluating representation for each observation task.

Detection timing and intervention outcomes require separate evaluation. In DriftGuard-SLA, earlier interventions accompanied an inconclusive reduction in violation duration relative to rules sharing its action mapping~\cite{driftguard2026}. This finding motivates assessing observation, intervention timing, and service outcomes separately. Our preliminary evidence covers annotated structural changes, scripted faults, and administrative retry limits. Their connection to service failure remains unestablished; conditional continuations have not been validated as forecasts.

\section{Future Plans}
\textbf{Ground the boundaries.} Relate candidate constraints to resource exhaustion and service outcomes under varied workload, using service-level indicators and error budgets~\cite{sre_alerts}. Derive limits on development histories, then evaluate consequences on unseen executions. Quantify uncertainty and distinguish a revised requirement from execution crossing an unchanged boundary.

\textbf{Maintain the runtime model.} Combine trace/configuration discovery with evidence of cross-version correspondence. Evaluate structural recovery and preservation of logical scope separately, including missing or conflicting metadata and changes of scope. Measure falsely accepted mappings, unavailable periods, update latency, and manual effort.

\textbf{Validate geometry and warning.} Compare raw-share and log-ratio continuations, smoothing, and reduced log-odds baselines on new histories. Hold telemetry and policy semantics constant, freeze rules before evaluation, and calibrate alarm burden independently. Evaluate actual first crossings and service-violation lead time. Decompose the roles of nonlinear representation, covariance estimation, and memory instead of attributing every difference to geometry.

\textbf{Evaluate decisions and control.} First compare diagnostic decisions against updated fraction/log-odds histories with identical counters, absolute volumes, workload history, and policy/change information. Then compare trigger policies using a common action set and mapping, including persistent threshold rules. Record both whether and when actions occur. Test whether trajectory-informed decisions reduce violations at comparable useful-work loss and intervention cost. Prespecified minimum useful effects and uncertainty bounds will distinguish evidence against the benefit hypothesis from an inconclusive estimate.

\section{Conclusion}
We propose a quantitative operationalization of the observable part of Rasmussen's drift model for evolving software. Recorded histories established comparable state assessment and distinguished movement of execution from revised constraints; a separate MEWMA experiment demonstrated representation-dependent detection. Further studies will test the operational value of trajectory-informed decisions through service-grounded boundaries and matched action comparisons.

\section*{Data Availability}
The implementation and replication materials are available in the research artifact~\cite{artifact}.

\clearpage
\bibliographystyle{IEEEtran}
\IEEEtriggeratref{19}
\bibliography{references}
\end{document}